\documentclass{article}

\usepackage{arxiv}

\usepackage[utf8]{inputenc} 
\usepackage[T1]{fontenc}    
\usepackage{hyperref}       
\usepackage{url}            
\usepackage{booktabs}       
\usepackage{amsfonts}       
\usepackage{nicefrac}       
\usepackage{microtype}      
\usepackage{lipsum}         
\usepackage{graphicx}
\usepackage{doi}

\usepackage{amsmath}
\usepackage{amssymb}
\usepackage{pifont}
\usepackage{paralist}
\usepackage{subcaption}
\usepackage{array,multirow}
\usepackage[table]{xcolor}
\newcolumntype{P}[1]{>{\centering\arraybackslash}p{#1}}
\newcommand{\cmark}{\ding{51}}%
\newcommand{\xmark}{\ding{55}}%

\title{Machine Learning Approaches for Active Queue Management: A Survey, Taxonomy, and Future Directions}

\newif\ifuniqueAffiliation
\uniqueAffiliationtrue

\author{ \href{}{\includegraphics[scale=0.00]{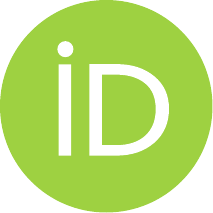}\hspace{1mm}Mohammad Parsa Toopchinezhad} \\
	Computer Engineering and \\Information Technology Department\\
	Razi University\\
	Kermanshah, Iran\\
	\texttt{ptoopchinzhad@gmail.com}\\
	\And
	\href{}{\includegraphics[scale=0.00]{orcid.pdf}\hspace{1mm}Mahmood Ahmadi} \\
	Computer Engineering and \\Information Technology Department\\
	Razi University\\
	Kermanshah, Iran\\
	\texttt{m.ahmadi@razi.ac.ir}\\
}

\renewcommand{\shorttitle}{Machine Learning Approaches for Active Queue Management}

\hypersetup{
pdftitle={Machine Learning Approaches for Active Queue Management: A Survey, Taxonomy, and Future Directions},
pdfsubject={AQM, ML},
pdfauthor={Mohammad Parsa Toopchinezhad, Mahmood Ahmadi},
pdfkeywords={First keyword, Second keyword, More},
}

\begin{document}
\maketitle

\begin{abstract}
Active Queue Management (AQM), a network-layer congestion control technique endorsed by the Internet Engineering Task Force (IETF), encourages routers to discard packets before the occurrence of buffer overflow. Traditional AQM techniques often employ heuristic approaches that require meticulous parameter adjustments, limiting their real-world applicability. In contrast, Machine Learning (ML) approaches offer highly adaptive, data-driven solutions custom to dynamic network conditions. Consequently, many researchers have adapted ML for AQM throughout the years, resulting in a wide variety of algorithms ranging from predicting congestion via supervised learning to discovering optimal packet-dropping policies with reinforcement learning. Despite these remarkable advancements, no previous work has compiled these methods in the form of a survey article. This paper presents the first thorough documentation and analysis of ML-based algorithms for AQM, in which the strengths and limitations of each proposed method are evaluated and compared. In addition, a novel taxonomy of ML approaches based on methodology is also established. The review is concluded by discussing unexplored research gaps and potential new directions for more robust ML-AQM methods.
\end{abstract}


\section{Introduction}
In 1986, the rapidly expanding yet still developing Internet encountered a significant new challenge known as ``congestion collapse'' \cite{gevros2001congestion}. This problem, caused by the inability of the network core to handle unregularized splurges of packets, led to a massive plummet in communication bandwidth—occasionally decreasing by 99.9\%. Fortunately, researchers were quick to develop an intuitive, yet highly effective set of mechanisms dubbed Congestion Control (CC). Designed to adaptively regulate data flow, CC effectively prevented total network-wide congestion, thereby "saving" the Internet \cite{jacobson1988congestion}. Since then, although modified from its original form, CC remains a cornerstone of contemporary packet-switched networks.

CC operates on the fundamental principle that by compelling individual nodes to behave less selfishly, all nodes will experience an improved service quality. More formally, it can be described as a distributed, network-wide algorithm aiming to intelligently manage data transmission to ensure stable and efficient communications. The necessity for CC emerges whenever a system is faced with demands exceeding its capacity \cite{floyd2000congestion}. Interestingly, CC is not confined to a single layer of the TCP/IP suite. Initially, strategies concentrated on the transport layer, but the concept of CC has spread to both the network and application layers. For instance, protocols like DASH, adjust the video's encoding quality based on congestion measurements along the path to the receiver. At the network layer, CC primarily takes the form of queue management \cite{peterson2022tcp}.

Historically, the responsibilities of the network layer were focused on forwarding and routing, and therefore it provided only basic queue management. For instance, when routers received multiple packets destined for the same location, they would simply queue the packets and process them using packet scheduling algorithms, typically First In First Out (FIFO) or Round Robin (RR) \cite{kurose2022computer}. Moreover, If the number of queuing packets exceeded the physical capacity or a software-defined one, routers would perform a simple action such as discarding the newly received packet (a strategy called Tail Dropping), discarding packets at the head of the queue (Head Dropping) or randomly discarding a packet (Random Dropping). Today, these crude queue management methods are known as Passive Queue Management (PQM) \cite{wei2017research}, and although straightforward, they served the early Internet well for many years.

\begin{figure}[!t]
\centering
\includegraphics[clip,width=\columnwidth]{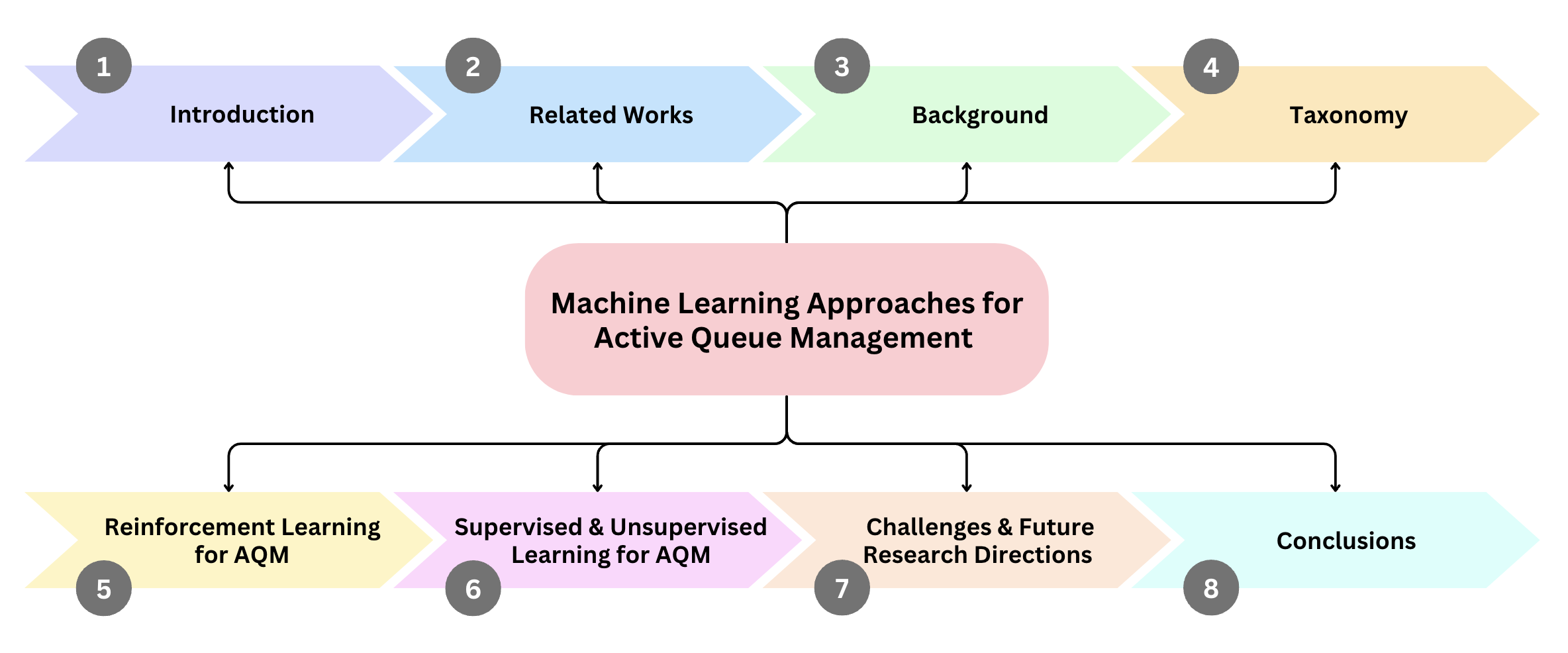}
\caption{The overall layout of this survey article.}
\label{fig_paper-structure}
\end{figure}

During the 90s, researchers began expressing concern that transport-layer CC, while effective, could not single-handedly advert the growing traffic of the Internet \cite{rfc2309}. An outdated belief that larger buffers would result in higher throughput and reduced packet loss has led to the design of oversized buffers. This practice had the unintended consequence of creating excessively large and frequently full queues, which ultimately increased Round-Trip Time (RTT)—a critical factor for modern low-latency applications such as video streaming and interactive gaming. This phenomenon, which would later become known as ``bufferbloat”, exposed the necessity of striking a balance between throughput and latency by carefully tuning router parameters \cite{gettys2012bufferbloat}. Given that this sort of management inherently could not be handled at the network’s edge, it became apparent that smarter control mechanisms had to be placed at the network core.

The solution to this problem was envisioned as a new class of queue management techniques known as Active Queue Management (AQM). These algorithms would reside in routers/switches and proactively drop packets before reaching full queue capacity. The rationale behind this approach was that the sender would interpret the dropped packet as an indication of high congestion levels, and adjust its sending rate before collapse had happened. It was theorized that AQM would effectively cut down on bufferbloat and provide low latency communications \cite{rfc2309}. In 1993, the first of these algorithms, named Random Early Detection (RED), \cite{floyd1993random} was published. Opting for a relatively simple strategy that would probabilistic drop packets before full capacity was reached, it was initially thought that RED would see large-scale adaptation among routers. However, RED never gained widespread use, primarily due to challenges associated with parameter tuning and an inability to adapt to altering network loads \cite{trinh2004comprehensive}.

To combat these shortcomings, other RED-inspired algorithms were proposed with the aim of autonomous, load-dependent parameter tuning \cite{adams2012active}. In addition to these heuristic approaches, control-theoretic and optimization-based methods were also thoroughly explored for their potential in queue management. Furthermore, newer concepts such as Explicit Congestion Notification (ECN) enabled AQM algorithms to mark packets instead of outright dropping them, significantly improving overall performance \cite{le2007effects}. However, despite these advancements, no single algorithm has been deemed robust enough to achieve widespread adoption as a standard practice \cite{rfc7567}. The emergence of 5G, the Internet of Things (IoT), and vehicular networks—along with their distinct characteristics and challenges—further emphasize that the quest for an effective AQM algorithm is still very much ongoing.

In parallel to these events, the field of Machine Learning (ML), a branch of computer science revolving around developing algorithms by leveraging data, had been properly maturing \cite{russell2016artificial}. Refined algorithms and dedicated libraries led to ML and its subsets of supervised, unsupervised, and reinforcement learning being implemented in every discipline imaginable \cite{lecun2015deep}. Although a limited number of papers had previously used ML for AQM, with the coming of this new age of ML, many new researchers started investigating the use of contemporary ML for AQM. The results of this trend led to many novel and creative works that employed state-of-the-art ML models to great success.

Despite the plethora of ML applications to AQM and their impressive results. No previous survey has been dedicated to the analysis of these works, even though one day, they very much might become the dominant form of AQM. This survey exists to bridge the gap between ML and AQM and to allow researchers from both sides of the spectrum to perceive even more powerful AQM algorithms. The main contributions of this paper are in the following:

\begin{enumerate}
   \item Establishing a novel taxonomy for ML approaches in AQM based on methodology.
   \item Documenting all major articles on supervised, unsupervised and reinforcement learning approaches for AQM.
   \item Detailing the shortcomings of current approaches and specifying promising future directions.
\end{enumerate}

The rest of this paper is organized as follows: Part II contrasts this work with preceding surveys. Part III provides a background on AQM, ML, DL, and its three subsets. In Part IV, a high-level view of the taxonomy for ML-based AQM is presented. Part V surveys reinforcement learning-based algorithms for AQM, while Part VI examines supervised and unsupervised approaches. Part VII provides a detailed list of current challenges and prospective directions of research, and finally, Part VIII concludes the paper. Figure \ref{fig_paper-structure} displays the overall structure of this paper and Table \ref{table_abrv} provides a list of used abbreviations.

\section{Related Works}
Numerous works surveying AQM methods exist in the literature. However, the majority of these reviews have become outdated or do not cover modern ML approaches. \cite{thiruchelvi2008survey} is one of the earliest articles examining AQM algorithms, it provides a basic categorization of methods as well as a brief overview of over 25 conventional methods, including the RED family of algorithms (SRED, DRED, ARED, etc). \cite{adams2012active} is another extensive survey covering AQM material from 1993 to 2011. This work provides an in-depth discussion of AQM schemes based on mechanism, context, and performance. Expectedly, as a work compiled before the rise of more advanced, deep learning methods, it catalogs only a few, basic neural network approaches that were limited in scale and power. \cite{li2014comparative} is a survey detailing neuron-based AQM controllers. This work pits four NN methods (Neuron PID, AN-AQM, FAPIDNN, and NRL), against three traditional ones (ARED, REM, and PI) and reports the overall superiority of neural-based methods in achieving higher accuracy, smaller queue length jitters, and better adaptability alongside faster convergence. However, in addition to only focusing on a small handful of methods, this decade-old survey predates all the exponential growth in ML methods, such as modern reinforcement learning and deep Q-learning. \cite{abbas2015fairness} is a study dedicated to fairness-driven queue management, although providing a useful resource on the topic, the paper practically does not mention any ML-AQM schemes. \cite{alwahab2018simulation} provides a simulation-based survey where 5 conventional methods (tail-drop, RED, PIE, CODEL, and PPV) were tested on heavy UDP and TCP traffic loads. Overall, the research found CODEL to perform best in end-to-end delay and jitter. 

Several ML-focused surveys about more general topics such as congestion control also exist. However, most of these works either briefly touch on AQM or fail to mention it entirely. \cite{jiang2021machine} is a general survey of ML methods for congestion control. However, due to the broad nature of this work, less than 8 papers on AQM are touched upon. \cite{zhang2020machine}, \cite{huang2021machine}, and \cite{wei2021congestion} are also examples of ML-CC surveys that take little to no notice of AQM.

In comparison, this article provides a thorough and focused examination of ML algorithms applied to AQM. We document the major contributions from its inception up to 2024, especially focusing on newer papers. Table \ref{table_comp}, provides an overview of this work compared to previous studies.

\begin{table}[!t]
\renewcommand{\arraystretch}{1.3}
\caption{List of commonly used abbreviations.}
\label{table_abrv}
\centering
\begin{tabular}{|c|c|}
\hline
Abbreviation & Description\\
\hline
CC & Congestion Control \\
IETF & Internet Engineering Task Force \\
QoS & Quality of Service \\
RFC & Request For Comments \\
AQM & Active Queue Management\\
RTT & Round Trip Time \\
RED & Random Early Detection \\
ECN & Explicit Congestion Notification \\
IoT & Internet of Things \\
ML & Machine Learning \\
DL & Deep Learning \\
NN & Neural Network \\
DNN & Deep Neural Network \\
RL & Reinforcement Learning \\
DRL & Deep Reinforcement Learning \\
MDP & Markov Decision Process \\
QL & Q-Learning \\
DQL & Deep Q-Learning \\
PPO & Proximal Policy Optimization \\
SDN & Software Defined Networking \\
\hline
\end{tabular}
\end{table}

\begin{table}[!t]
\renewcommand{\arraystretch}{1.3}
\caption{A comparison between this paper and exisitng surveys.}
\label{table_comp}
\centering
\begin{tabular}{|c||c|c|c|}
\hline
Reference & Year & ML-focused & AQM-focused \\
\hline
\hline
\cite{thiruchelvi2008survey} & 2008 & \xmark & \cmark \\
\hline
\cite{adams2012active} & 2012 & \xmark & \cmark \\
\hline
\cite{li2014comparative} & 2014 & $\thicksim$ & \cmark \\
\hline
\cite{abbas2015fairness} & 2015 & \xmark & \cmark \\
\hline
\cite{alwahab2018simulation} & 2018 & \xmark & \cmark \\
\hline
\cite{jiang2021machine} & 2021 & \cmark & \xmark \\
\hline
This Paper & 2024 & \cmark & \cmark \\
\hline
\end{tabular}
\end{table}

\section{Background}
This section elaborates on the history and development of AQM, summerised in Table \ref{table_rfc}, and also provides an outline of modern machine learning, deep learning, and its three major subsets.
\subsection{Active Queue Management}

\begin{table}[!t]
\renewcommand{\arraystretch}{1.3}
\caption{Summary of RFCs directly regarding AQM.}
\label{table_rfc}
\centering
\begin{tabular}{|c|c|p{7cm}|}
\hline
Name & Year & Relevant points \\
\hline
RFC 2309 & 1998 &  \begin{compactitem} \itemindent=-13pt \item Advocation of AQM in routers \item Recommendation of RED algorithm \end{compactitem} \\
\hline
RFC 7567 & 2015 &  \begin{compactitem} \itemindent=-13pt \item The explicit obsoletion of RED \item Emphasis on self-tuning algorithms \item Suggestion of ECN for packet marking \end{compactitem} \\
\hline
RFC 7928 & 2016 &  \begin{compactitem} \itemindent=-13pt \item Providing end-to-end evaluation metrics \item Specification of testbed topologies \end{compactitem} \\
\hline
RFC 8033 & 2017 &  \begin{compactitem} \itemindent=-13pt \item Proposal of PIE \end{compactitem} \\
\hline
RFC 8034 & 2017 &  \begin{compactitem} \itemindent=-13pt \item Outlining AQM demands for DOCSIS \item Proposal of PIE-DOCSIS \end{compactitem} \\
\hline
RFC 8289 & 2018 &  \begin{compactitem} \itemindent=-13pt \item Detailing CoDel \end{compactitem} \\
\hline
RFC 9332 & 2023 &  \begin{compactitem} \itemindent=-13pt \item Highlighting the role of AQM for L4S \item Specification of DualQ framework \end{compactitem} \\
\hline
\end{tabular}
\end{table}

\begin{figure}[!t]
\centering
\includegraphics[clip,width=11cm]{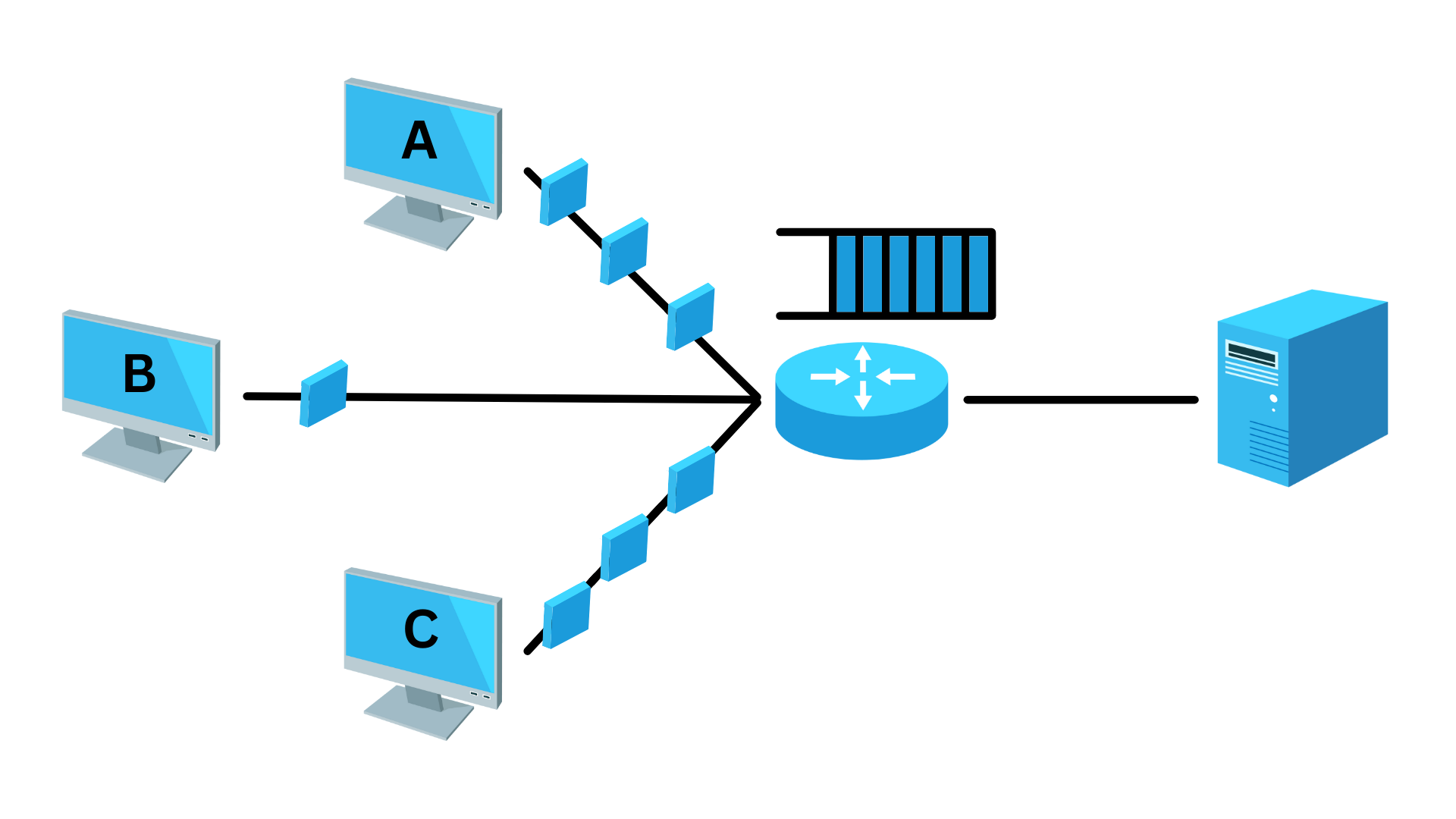}
\caption{Lock-out: The elephant flows running from sources A and C have complete hegemony of the router’s resources, as a result, source B’s packets will be consistently dropped.}
\label{fig_lock-out}
\end{figure}

RFC 2309 \cite{rfc2309} was the first IETF publication to formally introduce AQM. It ``strongly'' called for the ``testing, standardization, and widespread deployment of AQM in routers''. This RFC admitted that while TCP congestion avoidance mechanisms were essential, they could not provide quality service across all scenarios, given the limitations inherent in edge node control. Overall, the critical role of AQM can be summarized by the following points:
\begin{enumerate}
   \item \textbf{Bufferbloat:} An unavoidable aspect of tail-dropping is that in congested periods, the network will consist of queues maintaining a practically full status at all times. Such a state will lower latency to a crawl. Furthermore, full queues struggle to accommodate bursts of packets, which are indeed more frequent throughout the Internet \cite{jiang2005internet}. As a result, when a burst of incoming packets encounters a full queue, all packets will be dropped, increasing packet loss and potentially throughput \cite{rfc2309}. AQM addresses this challenge by reducing queue sizes while ensuring sufficient capacity to absorb and transmit packets during quieter periods.
  \item \textbf{Lock-Out:} A tail-dropping strategy tends to favor long-lasting flows (referred to as ``elephant flows'') over shorter, burstier flows (known as ``mice flows'') \cite{guo2001war}. This monopolization occurs since a periodically emitted packet is more likely to encounter a full queue (and consequently be dropped) than a stream of consistent packets (Figure \ref{fig_lock-out}). This issue can worsen when specific arrival patterns occur. AQM averts this potential problem by always reserving some space in the queue.
      \item \textbf{Aggressive Flows:} TCP-CC protocols depend on edge nodes to implement congestion management to reduce transmission rates during periods of congestion. However, not all data flows conform to this model. Unresponsive flows, particularly those generated by certain UDP-based applications such as streaming video and voice, often lack effective congestion avoidance mechanisms. If left unregulated, these unresponsive flows can lead to severe congestion collapse. Additionally, some flows, whether due to intentional design choices or faulty implementations, operate in a manner that is incompatible with CC, allowing them to unfairly monopolize network bandwidth. Effectively managing these problematic flows is ideally addressed at the networking layer because it provides a holistic approach to congestion management, ensuring fairness and stability across all types of network traffic \cite{rfc2309}.
  
\end{enumerate}

RFC 2309 recommended RED as an effective AQM algorithm to solve the above predicaments. RED \cite{floyd1993random} provides a straightforward solution to the queue management dilemma. A minimum ($min_\mathrm{th}$) and maximum ($max_\mathrm{th}$) threshold are determined. In addition, a sliding window average of the queue length is also kept ($q$). Upon arrival of a new packet, if the current average queue length is below $min_\mathrm{th}$, the packet is queued; if it exceeds $max_\mathrm{th}$, it is discarded. Yet, if the average falls between $min_\mathrm{th}$ and $max_\mathrm{th}$, the packet is dropped probabilistically based on $max_p$. Altogether, the drop probability ($d_p$) is defined as:

\begin{equation}
\label{eq_red}
d_p(q) = 
\left\{\begin{matrix}
0 & q < min_\mathrm{th} \\
max_p \frac{q - min_\mathrm{th}}{max_\mathrm{th} - min_\mathrm{th}}  &  min_\mathrm{th} < q < max_\mathrm{th} \\
1 & q > max_\mathrm{th} \\
\end{matrix}\right.
\end{equation}

Following the introduction of RED, many researchers began developing new AQM methods. The works in this period diverged into two principle directions: heuristic algorithms inspired by RED and control-theoretic approaches that applied mathematical controllers. 

Heuristic approaches can be broadly broken up into two. Some focused on enhancing the original RED algorithm by increasing steadiness like Stabilized RED \cite{ott1999sred} or by adaptively tuning its various parameters for higher responsiveness like Adaptive RED \cite{floyd2001adaptive}. Others developed novel algorithms that, although influenced by RED, would use different congestion indicators and dropping rules to mitigate delay and maximize utilization. The BLUE algorithm \cite{feng2002blue} utilizes link load as a source of congestion. After a fixed time interval, the drop probability would increase by a small fixed increment if any packet loss occurred and decrease if the queue became empty. Other algorithms such as Yellow \cite{long2005yellow}, BLACK \cite{chatranon2003black}, and GREEN \cite{feng2002green} also fall into this category. A detailed analysis of heuristic techniques can be found at \cite{thiruchelvi2008survey}.

Control-theoretic approaches are based on control theory, a developed field of applied mathematics and engineering, which provides a framework for autonomously controlling dynamic systems in achieving a desired state \cite{doyle2013feedback}. Control theory techniques generally rely on modeling a feedback control system and then applying a controller to adjust the control signal, e.g., drop probability, to attain a predefined input reference such as desired queue length. Various schemes incorporating PI \cite{sun2012iapi} or PID controllers \cite{yanfie2003design} for the task of AQM have been proposed, with others applying fuzzy logic controllers \cite{fengyuan2002design} or metaheuristic algorithms for parameter tuning \cite{chen2009ga}.

Although not yet widespread, the first ML approaches appeared in the literature during this period. Typically, these works would use basic to tune CT controller parameters \cite{sun2006neuron} or utilize the predictive abilities of NNs for early congestion prediction \cite{hariri2007nn}. The lack of sophisticated deep learning models and robust RL made these initial ML approaches limited in scale and applicability. A comparative survey on these early ML algorithms exists at \cite{li2014comparative}.

RFC 7567 \cite{rfc7567} was published 15 years after RFC 2309 and was intended to update AQM recommendations to reflect the ``experience and new research'' gained throughout the years. It begins by acknowledging the heightened significance of low-latency communication on the Internet, further increasing the importance of effective AQM. It then goes on to declare RED, due to difficulties in dynamic parameter adjusting, an obsolete and failed algorithm. Instead of proclaiming a new successor to RED, RFC 7567 advocates a set of recommendations for selecting an appropriate AQM algorithm. The two most important of which can be summarised as follows:

\begin{enumerate}

\item \textbf{Importance of self-tuning:}
The IETF recognized the rigidity of heuristic algorithms to be their most limiting factor. Therefore, it condemned any method that relies on manual parameter adjustment. Specifically, a candid AQM must provide a default behavior that auto-tunes to "operational conditions," such as the current observation of queue size, experienced delay, and packet loss.

\item \textbf{Support of packet marking:}
Defined in RFC 3168 \cite{rfc3168}, ECN is an extension to the TCP/IP stack that aims to provide better CC. It works by adding two bits to the IP header. Routers are allowed to change the sequence of these bits to convey impending congestion to the sender--which can adjust its transmission rate accordingly. ECN enables AQM algorithms to "warn" clients in times of moderate congestion by marking their packets instead of dropping them. This simple strategy can improve throughput and packet loss rate \cite{le2007effects}.

\end{enumerate}

\begin{figure}[!t]
\centering
\includegraphics[clip,width=11cm]{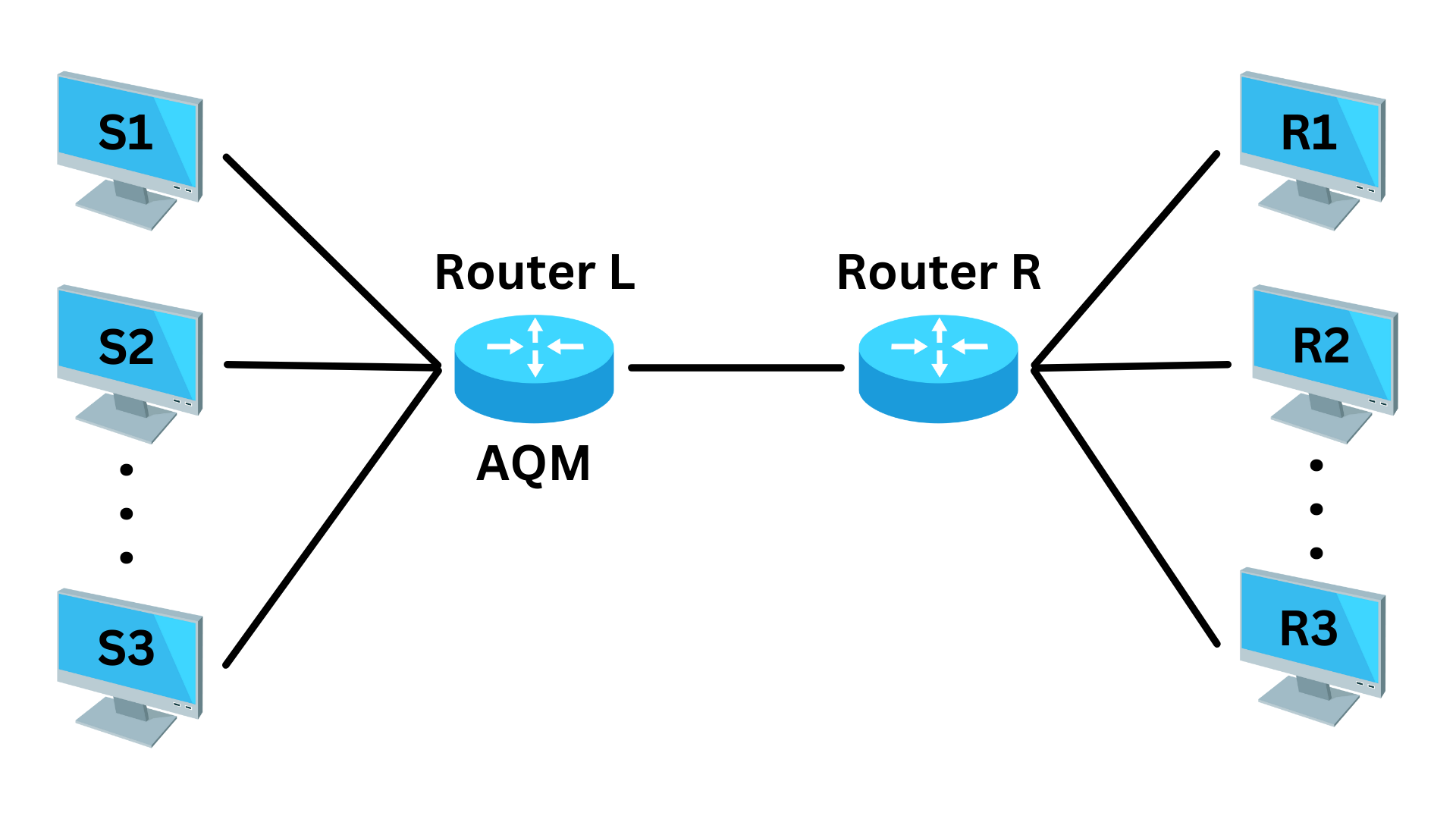}
\caption{The bottleneck or dumbell network topology, which is the IETF recommended testbed for AQM algorithms. The tested AQM algorithm should be placed at the left router.}
\label{fig_bottleneck}
\end{figure}

The next chronological IETF publication is RFC 7928 \cite{rfc7928}. It is a comprehensive work providing detailed guidelines for appropriately testing a proposed AQM scheme. It begins by listing six end-to-end metrics for accurately verifying AQM performance. Furthermore, it specifies a generic network topology, the bottleneck (Figure \ref{fig_bottleneck}), as a testbed. This work also proposes a series of unique scenarios to better evaluate fairness, stability, and more.

RFC 8033 \cite{rfc8033} presents Proportional Integral Controller Enhanced (PIE), a control theory-inspired AQM design. On the surface, PIE acts similarly to RED as it randomly drops incoming packets at the outbreak of congestion. However, under the hood, PIE uses a different form of congestion detection, queuing latency as opposed to queue length, to distinguish potential congestion. PIE utilizes the derivative of queuing latency as a measure to better understand traffic trends and maintain its latency target. A full description of PIE can be found at \cite{pan2013pie}.

RFC 8033 was picked off by RFC 8034 \cite{rfc8034}. This article discusses a practical implementation of PIE for Data-Over-Cable Service Interface Specification (DOCSIS)-based modems, named DOCSIS-PIE. This algorithm is specifically designed to combat cable modem bottlenecks, which are often culprits of high delays due to bufferbloat.

Detailed in RFC 8289 \cite{rfc8289}, Controlled Delay (CoDel) is one of the more successful AQM algorithms that has seen real-world adoption and is also implemented in the Linux kernel. CoDel distinguishes between so-called "good queues," which absorb and transmit bursts without the buildup of congestion, and "bad queues" which are persistently full. It is the former queue type that leads to bufferbloat, therefore CoDel only aims to avoid "bad queues." CoDel accomplishes this task by analyzing sojourn time, defined as the amount of queueing delay experienced by a packet. A well-behaved queue will frequently drain to zero and therefore many packets will experience near-zero sojourn times. In contrast, packets inside a congested router will consistently have large queueing delays \cite{peterson2022tcp}. CoDel uses this simple mechanism to identify and clear bufferbloat.

RFC 9332 \cite{rfc9332} is the most recent document regarding AQM. This RFC describes a Dual Queue (DualQ) system for integrating AQM algorithms into two separate queues, each designated for a specific flow. The main incentive for this framework is to support the newly envisioned Low Latency, Low Loss, and Scalable Throughput (L4S) architecture \cite{rfc9330}. L4S is a profound topic, but in short, it is a novel congestion control mechanism designed for real-time applications to reduce network latency to an average of 1 millisecond while keeping high throughputs. DualQ classifies all incoming packets into those sent with scalable congestion controls (e.g., DCTCP, TCP Prague) and ’Classic’ flows (e.g., Reno, CUBIC). This way, a separate queue and AQM algorithm is designated for each kind of flow \cite{de2022dual}.

\subsection{Machine Learning}
ML is a subset of Artificial Intelligence (AI) that focuses on algorithms that analyze swaths of data to extract meaningful patterns, many of which are imperceptible to the human eye \cite{jordan2015machine}. ML can be broadly grouped into three major subsets: supervised learning, unsupervised learning, and reinforcement learning. While ML can trace its roots to the 1950s, it is only in recent decades that an abundance of computing power and datasets, coupled with refined algorithms, has propelled ML into a prominent position in both research and industrial applications.

The successful applications of ML are numerous and varied. These range from the development of general-purpose large language models (LLMs) used in chatbots \cite{zhao2023survey}, to autonomous vehicles \cite{bachute2021autonomous}, face recognition systems \cite{wang2021deep} and so on. Many of the most impressive feats of ML emerge from its integration with other fields, leading to exciting interdisciplinary work in areas such as healthcare \cite{qayyum2020secure} (e.g., diagnosis and treatment recommendations), agriculture \cite{meshram2021machine} (e.g., precision farming and crop monitoring), and finance \cite{dixon2020machine} (e.g., algorithmic trading and fraud detection).

The utilization of ML in networking has also experienced significant growth in recent years \cite{kanakis2022machine}. This surge, along with advancements in IoT, 5G technology, and Software-Defined Networking (SDN), has given rise to concepts like "Artificial Intelligence of Things" (AIoT) \cite{zhang2020empowering} and Knowledge-Defined Networking (KDN) \cite{mestres2017knowledge}. ML-based solutions are now being explored for nearly every layer of the Internet stack, addressing challenges in areas such as network routing \cite{mammeri2019reinforcement}, network security \cite{uprety2020reinforcement}, IoT task offloading \cite{zabihi2023reinforcement}, traffic classification \cite{pacheco2018towards}, caching \cite{zhu2018deep} and much more \cite{boutaba2018comprehensive}. These methods have repeatedly shown better performance while providing adaptability and generalizability to varying network conditions.

\subsection{Deep Learning}

Much of the success of ML can be attributed to Deep Learning (DL). Traditional ML algorithms such as decision trees, support vector machines, and neural networks were generally unable to process high-dimensional raw data. This limitation would require researchers and engineers to meticulously handcraft input features. Starting from the early 2010s, methods were developed, typically using multilayer neural networks in opposition to the shallow neural networks that contained only a single hidden layer. These new models proved to be powerful learning models capable of extracting raveled patterns in high-dimensional data, which resulted in the breaking of many historical records in fields like image or speech recognition. These newly called Deep Neural Networks (DNN) led the spearhead of the DL revolution and began to be quickly used in all fields of science and engineering as the need for carefully picked features was removed \cite{lecun2015deep}.

A hallmark characteristic of DNNs is their ability to adapt to various data types and tasks. For image processing tasks, Convolutional Neural Networks (CNNs) have been developed that learn filters to better extract visual data \cite{o2015introduction}. Recurrent Neural Networks (RNN) were developed to enable NNs to have a form of memory, turning them into appropriate tools for processing sequential data \cite{yu2019review}. Today, DNNs have made their way into state-of-the-art supervised, unsupervised, and reinforcement learning algorithms.

\subsection{Supervised Learning}
The SL problem consists of training a model to correctly map input features to target outputs. SL methods rely on the existence of a labeled dataset that provides the correct labels for each entry \cite{jordan2015machine}. Generally, datasets are split into two parts, with one being used to train the model and the other used for testing the accuracy of its learned knowledge. Due to its applicable nature, SL is the most researched and applied subset of ML and has been widely adopted in the real world.

SL problems are generally classified into two primary categories: classification and regression. In classification, the model is trained and subsequently used to predict discrete labels such as type of malware \cite{milosevic2017machine}. Regression, on the other hand, focuses on predicting continuous numerical values, such as the predicted network traffic volume \cite{ramakrishnan2018network}.

Despite its typically straightforward and effective nature, SL comes with significant limitations. The most prominent is the requirement for a sufficiently large labeled dataset, where the involvement of professionals is often essential for accurate labeling. In the realm of computer networking research, SL has demonstrated impressive performance when working with clean, large-scale datasets, such as intrusion detection \cite{sharafaldin2018toward}. However, applying SL to domains lacking adequate labeled datasets remains a considerable challenge.

\subsection{Unsupervised Learning}
In the USL framework, algorithms are applied to uncover patterns and structures in unlabelled data. This approach eliminates the expensive procedure of labeled-data collection, as excessive quantities of unlabeled data naturally exist across various domains, e.g., large corpora of text scraped from the Internet or routine logs generated from networking devices.

USL encompasses tasks like clustering (the grouping of similar data), dimensionality reduction (the shrinking of datasets by decreasing the number of features), and anomaly detection (the identification of irregular data). In the context of computer networking, USL has found applications in traffic pattern recognition and data flow monitoring \cite{usama2019unsupervised}.

\subsection{Reinforcement Learning}
RL concentrates on training decision-making agents to maximize rewards through trial-and-error interactions within a given environment \cite{arulkumaran2017deep}. Its robust problem-solving properties and lack of reliance on predefined datasets—unlike other ML subsets—have enabled its successful application across various domains. Because many computer networking problems can be naturally framed as sequential decision-making processes, RL has gained significant traction in this field \cite{boutaba2018comprehensive}. These properties position RL at the forefront of ML-based approaches for AQM. Thus, we will conduct a comprehensive analysis of RL in this section.

In an RL system, each timestep consists of the agent perceiving its environment through an observation, deciding on an action,  and subsequently receiving feedback regarding the outcome of that choice. This entire process is formalized using a mathematical framework called a Markov Decision Process (MDP). An MDP is represented as a tuple $\langle S, A, T, R, \gamma \rangle$, where $S$ is the set of all possible states, $A$ is the set of available actions, $T$ maps a state-action pair to a probability distribution of the next state, $R$ represents the reward function which determines the reward for taking an action in a specific state and $\gamma$ is a discount factor that determines the importance of future rewards relative to immediate ones \cite{fenjiro2018deep}. The objective of RL algorithms is to discover an optimal mapping of states to actions (known as a policy) that maximizes the overall discounted reward received over time. To this end, RL agents must intelligently balance between exploring the environment (by taking unknown yet potentially rewarding actions) and exploiting it (capitalizing on actions that have proven successful in the past) \cite{ladosz2022exploration}.

The tipping point for RL came around 2013, when researchers successfully adapted DNN with traditional RL methods, ushering in the deep reinforcement learning era \cite{mnih2013playing}. Before this innovation, RL algorithms primarily relied on tabular methods for storing state or action values. DRL replaced these rudimentary tables with state-of-the-art, multi-layered DNN approximators, enabling the capture of high-dimensional features. Furthermore, DNNs offer enhanced generalization capabilities, allowing them to efficiently represent vast state spaces.

\begin{figure}[!t]
\centering
\includegraphics[clip,width=13cm]{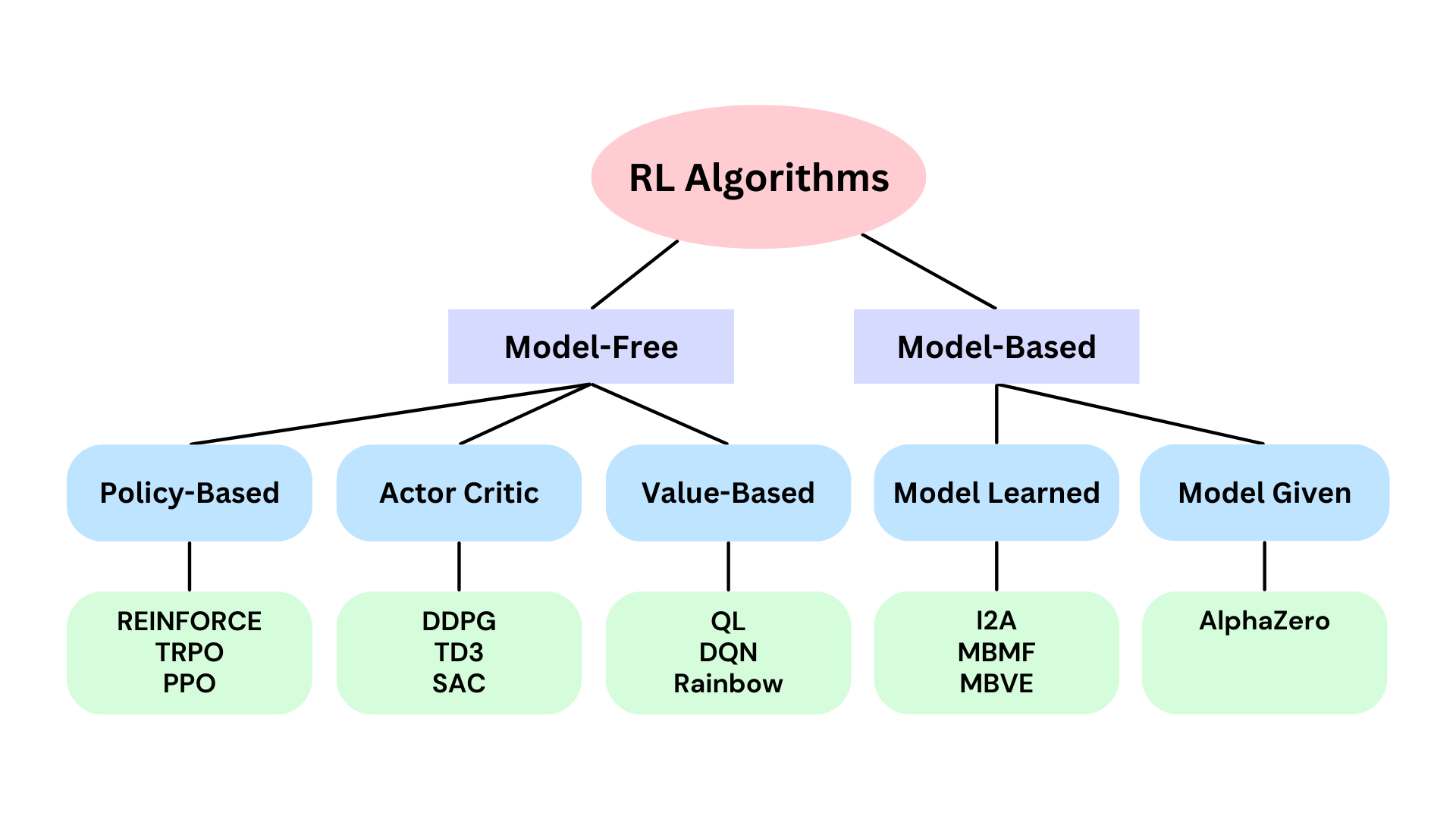}
\caption{A non-exhaustive taxonomy of RL algorithms.}
\label{fig_rltax}
\end{figure}

\begin{figure*}[!t]
\centering
\includegraphics[width=\columnwidth]{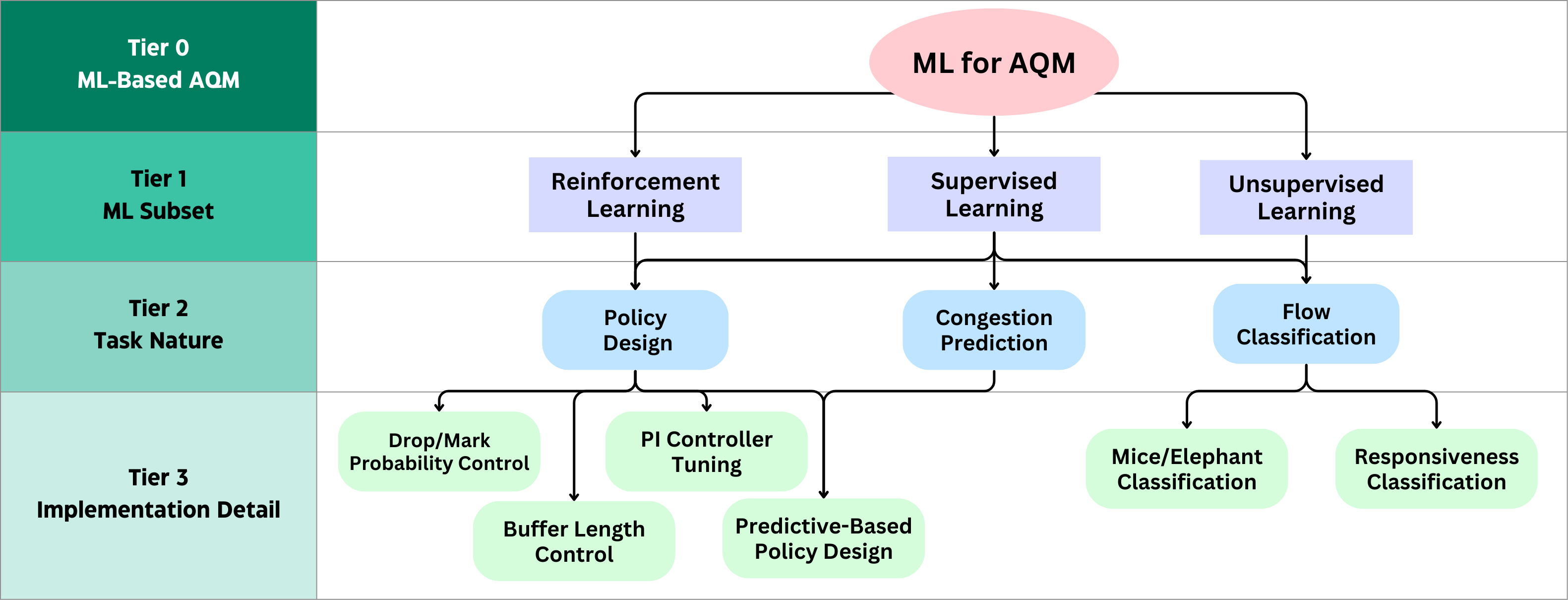}
\caption{A taxonomy of ML approaches for AQM. The first tier separates approaches based on applied ML techniques while the second and third tiers categorize by the specific application implementation detail, respectively.}
\label{fig_aqmtax}
\end{figure*}

Figure \ref{fig_rltax} illustrates the landscape of modern RL \cite{zhang2020taxonomy}. Algorithms either fall into the model-based or model-free category, depending on whether they utilize a predictive model. A model is defined as a function that determines the next state based on the current state and action: $M(S,A) = S'$. Model-based algorithms can be provided an accurate model before training or attempt to construct one by training on the data it gathers. While model-based methods can be sample efficient, they have notoriously been more challenging and unstable to train \cite{moerland2023model}.

Model-free algorithms are by far the most applied and thoroughly researched RL methods. The principal goal of any RL algorithm is to derive an optimal policy. Some methods find such a policy directly (policy-based), while others focus on estimating state or action-state values first (value-based), and some combine both approaches (actor-critic). In policy-based methods, the policy is typically represented as a parameterized function, e.g., a neural network. Optimization techniques are then employed to find the set of parameters that maximizes the policy's effectiveness, as determined by the reward function. Proximal policy optimization (PPO) \cite{schulman2017proximal} is widely recognized as the current state-of-the-art policy-based algorithm. 

In value-based RL, algorithms focus on estimating the expected cumulative future rewards associated with being in a specific state while following a selected action. Deriving the policy after establishing the Q function is a straightforward task, effectively assessing the quality of a state-action pair. This assessment is typically represented by the Q-function. Actor-critic models, on the other hand, take a mixed policy and value-based approach to problem-solving. These methods consist of two components: one that chooses the action and another that evaluates the quality of the action taken \cite{arulkumaran2017deep}. It is worth noting that the line between actor-critic and some policy-based methods, like PPO, is fuzzy since they utilize a form of value-function, but for simplicity's sake, we specify these algorithms under the policy-based category.

\section{Proposed Taxonomy}

As the encompassing scope of this survey, only original research papers that explicitly enhance AQM via a form of ML are considered. Figure \ref{fig_aqmtax} depicts the proposed taxonomy. Tier 1 classifies approaches according to the applied subset of machine learning: supervised learning, unsupervised learning, and reinforcement learning. This classification is grounded in the fundamental differences among these subsets, as each addresses problems unique to their capabilities and formulation. Moreover, organizing by ML subset is particularly relevant, given that the majority of surveyed papers fall neatly within one of these subsets.

The second tier categorizes methods based on the objective of the ML model, i.e., the AQM issue that the model aims to address through its application. Here, three primary categories become prevalent in the literature. These are policy design, congestion prediction, and flow classification.

\subsection{Policy Design}
Policy design is the most natural and direct framework to apply ML to AQM. As a result, it is also the most prevalent in the literature. The papers in this category directly control the queue to ensure high network performance. Policy design is feasible with both SL and RL, with the former formulating AQM as a general optimization problem \cite{sun2007adaptive} and the latter rewriting AQM as an MDP.\cite{kim2021deep}.

Policy design approaches can be further broken down by account of their precise form of control. Drop/mark probability control methods opt to directly control the probability in which packets are dropped or marked \cite{ma2021intelligent}. Buffer length control techniques dynamically adjust the maximum queue length by increasing or shrinking the queue size \cite{bouacida2018practical}. Controller tuning-focused strategies adopt ML models to tune the parameters of conventional PI or PID controllers \cite{sun2006neuron}, typically resulting in an architecture known as a neuron controller \cite{fang2011design}.

\subsection{Congestion Predection}
Congestion predictors utilize the powerful forecasting capabilities of ML algorithms to predict the future state of network load \cite{sneha2020prediction}. A traditional AQM algorithm can then use this information to govern queues with anticipation of the near future. Congestion control is a well-studied topic in computer networking \cite{mohammed2019machine}. For our purposes, however, only works utilizing it within the context of AQM are considered. The earliest articles in this category employed basic neural networks \cite{hariri2007nn}. More recent research uses more powerful RNN architectures, such as Long Short-Term Memory (LSTM) \cite{hu2019nonlinear}, which better exploit the temporal dependencies across sequential data \cite{yu2019review}. Another advantage of this method is that congestion data collection is a typically straightforward process as the network self-generates this data without the need for any human intervention, making it an ideal approach for real-world deployment. 

Predicted congestion levels can also be incorporated into the observation of RL algorithms to underpin decision-making \cite{gomez2021federated}. This hybrid SL and RL approach leads to the predictive-based policy design category, blending the benefits of both.

\subsection{Flow classification}
Flow classification involves dividing different types of flows based on their observed features. These segregated flows can then be placed into different queues, each governed by a designated AQM algorithm. Flow classification methods range from simple heuristic-based to more complex ML-based methods. Supervised learning models are a popular choice for classification problems. These models are trained on labeled datasets containing many rows of historical data. While obtaining a refined, large dataset can be gruelling, the benefits are often worthwhile, as accurately separating flows has been shown to increase various performance criteria \cite{taylor2005survey}.

Unsupervised learning is also capable of successfully classifying flows by employing techniques such as outlier detection or clustering based on extracted features \cite{hochst2017unsupervised}. While implementing unsupervised learning for flow classification can be more precarious compared to a straightforward supervised learning approach, the absence of labeled data makes it a valuable alternative worth considering. Unsupervised learning can even be paired with supervised learning to create robust two-phased classification schemes \cite{bakhshi2016internet}.

As for the basis of classification, many different criteria exist. In the context of AQM, two predominant themes emerge: the differentiation between ``mice" and ``elephant" flows, and the distinction between responsive and unresponsive flows.

As stated in Section III, a primary motivation for implementing AQM is to avoid "lockout," wherein dominant elephant flows monopolize a router's resources, leaving little to no available space for smaller mice flows. By effectively segregating these flow types, one can alleviate the lockout issue and promote a more equitable distribution of network resources \cite{majidi2020dc}.

The classification of responsive versus nonresponsive flows is crucial for solving the aggressive traffic problem. AQM algorithms can leverage this classification to allocate more resources to compliant, responsive flows while punishing nonresponsive ones \cite{latre2013cognitive}.

\section{Reinforcement Learning for AQM}

\begin{figure*}[!t]
\centering
\includegraphics[width=11cm]{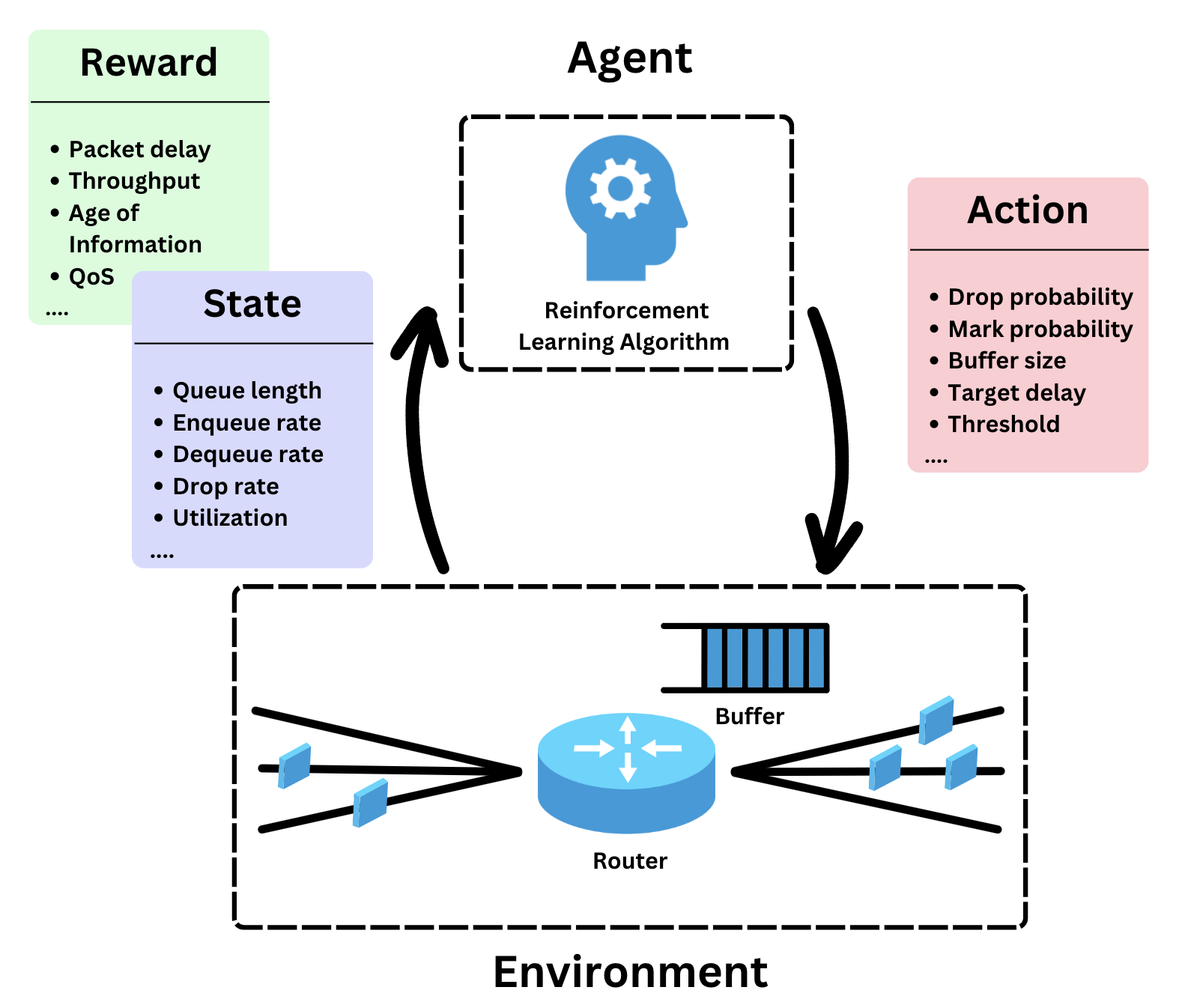}
\caption{The AQM problem formulated as an MDP, showcasing the most common state space, action space, and reward functions proposed in the literature.}
\label{fig_rl-scheme}
\end{figure*}

RL techniques are by far the most applied ML subset for AQM. As a result, a large body of innovative RL schemes exist in the literature, all of which fall into the policy design branch. These approaches generally have the same premise: they begin by defining the properties of the MDP and then employ RL algorithms to search for effective policies (Figure \ref{fig_rl-scheme}). Table \ref{table_rl-aqm} presents a list of the surveyed RL schemes, summarizing the main distinction between the works such as the employed RL algorithm, applied network scenario, and MDP formulation. The list is organized based on the control strategy mechanism, as depicted in tier three of Figure \ref{fig_aqmtax}. The identified categories consist of drop probability control, marking probability control, buffer size control, and a miscellaneous category for unique approaches.

\subsection{Drop Probability Control}
The methods in this category can be viewed as the successors of the original RED algorithm. As discussed in Section III, two of RED's shortcomings are its reliance on manual parameter tuning and its inability to adapt to changing network conditions. These drawbacks were identified early on, prompting researchers to employ the prominent techniques of the time, such as fuzzy control, to dynamically adjust parameters \cite{sun2007stabilizing}. The subsequent algorithms presented in this section build on this foundation by harnessing the data-driven decision-making capabilities of RL. Furthermore, the more recent approaches incorporate DRL, which allows for even more robust pattern identification across a much broader state space.

QRED \cite{su2018qred} represents the simplest manifestation of RL for drop probability control. It substitutes RED's heuristic-based decision-making method, Eq. \eqref{eq_red}, in place of a tabular Q-Learning algorithm. QRED utilizes the same data as RED (average queue length) as its observation, but rather than managing three parameters ($min_\mathrm{th}$, $max_\mathrm{th}$, $max_p$), it focuses solely on controlling drop probability to promote quicker learning. The reward function is defined as a simple weighted sum of the average throughput and inverse of delay. Despite QRED's straightforwardness, its efficiency is demonstrated through network simulations. Both RED and QRED were tested across low, medium, high, and variable loads within a standard bottleneck topology. QRED consistently achieves higher throughput, lower delay, and reduced packet loss in nearly all scenarios, with the difference becoming more pronounced as the load increases. This work was among the first to illustrate the potential of RL for AQM.

RLAQM \cite{jiang2020performance} is conceptually similar to QRED. A key distinction is the addition of the differentiated average queue length into the state space. Furthermore, while QRED manages drop probability, RLAQM controls the $max_p$ parameter. This method's most notable drawback is that it does not control the other two parameters of RED which limits it ability to adapt under specific load conditions. Nevertheless, RLAQM still demonstrates superior performance compared to RED.

The methods proposed in \cite{alwahab2021deep} and \cite{basheer2021intelligent} were among the first to leverage the mightier DQL framework. Utilizing a DNN approximator instead of a table provided these techniques with increased generalization and also the ability to accommodate a larger state space that incorporates factors such as link utilization and delay. This provides the RL agent with a richer source of information for decision-making. The key difference between these two methods lies in their action selection strategies. \cite{alwahab2021deep} adjusts the drop probability by either 0.1, 0.01, or 0 per action, while \cite{basheer2021intelligent} selects a dropping probability from a continuous range of 0 to 1. The flexibility of the former is limited, as adjusting by a fixed value can result in slower reactions to extreme changes and an inability to fine-tune probabilities for precise control.

Proposed in \cite{ma2021intelligent}, DRL-AQM presents a series of enhancements to the RL-AQM scheme. A key innovation is how it integrates historical router data into the state space. Concretely, the agent's observations incorporate the preceding ten values of metrics such as queue length, send rate, and dequeue rate. This capability allows the system to retain information from past trends and is a common practice in RL \cite{mnih2015human}. Furthermore, DRL-AQM uses a state-of-the-art policy iteration algorithm, PPO, instead of the prevailing value iteration DQL algorithm due to its better performance. This paper also concluded that for the task of selecting max dropping probability, a continuous action space outperforms a discrete one.

\cite{kim2021deep} offers a fresh perspective on RL-based AQM, with its focus on an IoT context. The simulated environment consists of various smart-home devices sending TCP or UDP messages via a home network to the cloud. The AQM agent utilizes a conventional DQL architecture and directly proposes drop probabilities, which are rewarded based on their effectiveness in maintaining a delay around a reference value while also lowering packet drop rates.

MACC (Multi-Agent Congestion Control) \cite{bai2022macc} investigates a cross-layer scheme combing traditional TCP-CC and AQM algorithms. This research employs a specialized variant of RL known as Multi-Agent Reinforcement Learning (MARL) \cite{zhang2021multi}. In MACC, one RL agent operates at the transmission layer to dynamically adjust the Congestion Window (CWND) and threshold values, while another agent functions at the network layer, managing the enqueue rate and packet drop rate. The overarching objective of this dual-agent system is to facilitate cooperation between the two agents to achieve higher throughput and lower latency compared to a single-agent scheme. MACC surpass mixtures of conventional CC and AQM algorithms (NewReno \cite{rfc3782}, BBR \cite{cardwell2016bbr}, CoDel) in both RTT and throughput.

\cite{ali2023efficient} leverages an ensemble of 4 DRL algorithms (DQN, PPO, DDPG, TD3) to perform AQM. The rationale behind this approach is that a single model often struggles to effectively navigate the expansive state and action spaces characteristic of many complex problems. Furthermore, a single RL agent can typically exhibit volatility and bias, which can detract from their practical applicability. By utilizing ensemble RL, multiple distinct RL algorithms are trained and their outputs are combined, allowing for a more comprehensive exploration of the problem space and improved generalization capabilities \cite{song2023ensemble}. In this paper, the authors design varying action spaces for each algorithm and compute the average of their outputs to determine the final selected action. The performance of this ensemble model is compared against both RED and the four individual RL algorithms and was found to achieve superior throughput. This paper showcases the computation-accuracy tradeoff and asks the question of whether multiplying computation costs are worth the benefits.

\subsection{Mark Probability Control}
The algorithms in this class are conceptually similar to those in the previous category but mark packets instead of dropping them. Whenever a packet is marked, the router allows the packet to propagate through the network and only warns clients to lower send rates to prevent a buildup of congestion. In its simplest form, an ECN algorithm has the same parameters as RED but substitutes dropping probability with marking probability. While it is true that the methods in these two categories are somewhat interchangeable, distinguishing between the two categories enables a more thorough analysis of each. This division is particularly relevant as ECN is more prevalent in specific types of networks like Data Center Networks (DCNs), which have distinct characteristics.

RLECN \cite{jung2023rlecn} proposes a Q-learning-based scheme where the RL agent dynamically adjusts the marking threshold of the system. The action space in this approach is limited as it has no way of changing the marking probability. A notable feature of RLECN is its integration of SDN. The SDN controller operates at a higher plane than the routers and can gather relevant information that allows for higher network visibility. RLECN controls congestion by shrinking or enlarging the marking threshold by 5, 10, or 0 packets at each timestep.

Queuepilot \cite{dery2023queuepilot} aims to balance high utilization with low loss rates and short delays. It automatically tunes the ECN marking probability after offline training on various settings, producing a lightweight policy that can be applied online without further learning. Experiments on real networks with hundreds of TCP connections showed that QueuePilot outperforms existing algorithms, including droptail and other advanced AQM schemes, especially in small buffer scenarios. Moreover, it demonstrates stable performance across a wide range of buffer sizes, adapting to varying levels of congestion and showing promise in handling complex topologies. The system is designed to be compatible with current router architectures, relying solely on local buffer information and avoiding the need for GPUs or per-flow tracking.

The following three works are all specifically designed for DCNs. Therefore, it is fitting to first examine these networks' general characteristics and unique requirements. DCNs serve as the backbone for large data centers, which are distinguished by high requirements of availability, scalability, and performance. Given that large data centers receive millions of daily requests, they can become significant communication bottlenecks. These properties have led to a multitude of studies proposing schemes for load-balancing and topology design to enhance data center flexibility \cite{xia2016survey}. One common structure for DCNs is the spine-leaf architecture, which is a two-tier system with a spine layer (containing high-capacity switches) and a leaf layer (consisting of access switches). This model allows for low hop count communications while still being cost-effective \cite{harsh2020spineless}. Evidently, Due to the high numbers of servers, links, switches, and routers typically found in a DCN, a centralized approach in which an agent observes the entire data center's state is not plausible. As a result, the papers in this category generally adopt a multi-agent approach.

ACC \cite{yan2021acc} was the first work to develop an RL-based AQM scheme for DCNs. ACC dynamically adjusts the three parameters of minimum threshold, maximum threshold, and marking probability. A DRL agent is placed in each switch of a leaf-spine network, forming an overall multi-agent system. Each DRL agent observes its local statistics and independently chooses an action for adjusting the ECN settings based on the reward function. ACC is compared with two static sets of ECN settings in various simulations and triumphs in both improving the flow completion time for smaller messages and maintaining high throughput for larger ones. Remarkably, ACC has also been deployed in a production high-speed data center with great success. The initial offline training phase of ACC's model plays a crucial role in its real-world applicability as training on realistic traffic traces before deployment significantly reduces the risk associated with the initial trial-and-error-filled agent decisions.

HAECN \cite{ma2022rlrbm} is a direct successor to ACC and addresses the high computational demands of ACC that complicate deployment on switches. To mitigate this issue, HAECN proposes a hierarchical RL-based ECN tuning algorithm. This advanced form of RL decomposes complex tasks into a hierarchical structure, wherein a higher-level controller issues overarching commands to a series of lower-level agents that handle specific subtasks \cite{pateria2021hierarchical}. Under this framework, HAECN splits the AQM algorithm into a policy and decision module. The policy module is activated only when the decision module detects that the current setting of ECN is not good enough and needs to be updated based on the current condition. This new design not only reduces the computation overhead but also achieves a 24\% higher throughput compared with ACC.

PET \cite{cheng2024pet}, another evolution of ACC, tries to relax the computational overhead by leveraging Independent Proximal Policy Optimization (IPPO) \cite{de2020independent} as its learning algorithm. This method lacks any need for global experience replay as each agent locally estimates its own value function and thus reduces both the overhead and bandwidth costs related to inter-agent information sharing. Moreover, PET embeds features of the traffic flows, especially the percentage of elephant versus mouse flows, into its observations. PET's training process follows the Distributed Time-Dependent Environment (DTDE) framework that consists of two phases: an offline training phase that uses historical network data collected by non-RL policies, followed by an online phase that fine-tunes the previously trained model to better adapt to specific network conditions. PET exhibits improved performance across a variety of network scenarios, achieving shorter flow completion times, enhanced model stability, faster convergence rates, and increased system robustness.

\begin{table*}[!t]
\caption{List of works applying RL to AQM}
\label{table_rl-aqm}

\def\arraystretch{1.02}
\hspace*{-1.5cm}
\begin{tabular}{| c |l| P{1.75cm} | P{1.75cm}  | P{5.5cm}  | P{2.5cm} | P{3cm} |}
\hline
\rowcolor{gray}
\bf{Type} & \multicolumn{1}{c|}{\bf{\textsc{Ref.}}} & \multicolumn{1}{c|}{\bf{\textsc{Algorithm}}} & \multicolumn{1}{c|}{\bf{\textsc{Network}}} & \multicolumn{1}{c|}{\bf{\textsc{State}}} & \multicolumn{1}{c|}{\bf{\textsc{Action}}} & \multicolumn{1}{c|}{\bf{\textsc{Reward}}}\\
\hline
\hline
\parbox[t]{2mm}{\multirow{18}{*}{\rotatebox[origin=c]{90}{Packet Dropping Control}}}
& \cite{su2018qred} & QL & Bottleneck & Avg. queue length & Drop probability & Max avg. throughput, min avg. delay\\
\cline{2-7}
& \cite{jiang2020performance} & QL & Bottleneck & Avg. queue length, $\Delta$avg. queue length & Max drop probability & Max throughput, min queue size\\
\cline{2-7}
& \cite{alwahab2021deep} & DQL & Bottleneck & Queuing delay, link utilization, drop probability & Increase/decrease drop probability & Max utilization, min queuing delay\\
\cline{2-7}
& \cite{basheer2021intelligent} & DQL & Bottleneck & Dequeue rate, enqueue rate, drop rate, avg. queue length & Drop probability & Max queuing delay, min drop rate \\
\cline{2-7}
& \cite{ma2021intelligent} & PPO & Bottleneck & Source send rate, number of dequeue packets, time slot gradient & Drop probability & Max utilization, min queue length, min drop rate\\
\cline{2-7}
& \cite{kim2021deep} & DQL & IoT (smart-home) & Queue length, dequeue rate, queuing delay & Drop probability & Max queuing delay, min drop rate \\
\cline{2-7}
& \cite{bai2022macc} & VDN & Bottleneck & Queue length, dequeue rate, queuing delay & Drop probability & Max throughput, min queue delay, min queue length\\
\cline{2-7}
& \cite{ali2023efficient} & DQL, PPO, DDPG, TD3 & Bottleneck & Queue length, number of dropped packets & Drop probability  & Max throughput, min drop ratio, min queue size\\
\cline{2-7}
\hline

\parbox[t]{2mm}{\multirow{16}{*}{\rotatebox[origin=c]{90}{Packet Marking Control}}}
& \cite{jung2023rlecn} & QL & Bottleneck + SDN controller & Number of incoming packets, number of outgoing packets, queue length & Increase/decrease mark threshold & Max utilization\\
\cline{2-7}
& \cite{dery2023queuepilot} & PPO & Bottleneck & Number of received packets, link utilization, packet drop rate, packet marking rate, avg. queue length, avg. queue delay & Mark probability & Max queuing delay, min drop rate\\
\cline{2-7}
& \cite{yan2021acc} & DQL & DCN & Queue length, output data rate for each link, output rate of ECN marked packets, ECN-settings & Max mark probability, max and min threshold  & Max utilization, min queue length\\
\cline{2-7}
& \cite{hu2023haecn} & DDQN, DQL & DCN & Queue length, link utilization, ECN-settings & Max mark probability, max and min threshold  & Max link utilization, min queue length\\
\cline{2-7}
& \cite{cheng2024pet}& PPO & DCN & Queue length, output rate, ECN-marked output rate, ECN threshold, incast degree, mice/elephant ratio  & Max mark probability, max and min threshold  & Max throughput, min queue length\\
\hline

\parbox[t]{2mm}{\multirow{10}{*}{\rotatebox[origin=c]{90}{Buffer Size Control}}}
& \cite{bouacida2018practical} & QL & Wireless & Current buffer size & Increase/decrease buffer size & High throughput, low delay\\
\cline{2-7}
& \cite{sneha2021bo} & DQL & Bottleneck & Number of dropped packets, avg. queue size, avg. packet delay, avg. throughput& Increase/decrease buffer size & Max throughput, min delay\\
\cline{2-7}
& \cite{ma2022rlrbm} & A3C & 5G (RAN) & RLC buffer content, DL-SCH capacity, data arrival rate,  congestion level, number of dropped packets& Maximum buffer size & Max throughput, min delay, min queue delay\\
\cline{2-7}
& \cite{song2024deep} & DQL & WSN & Channel conditions, current time, AoI value, frame timestamp, in-queue location & Forward, flush or maintain buffer & Min AoI values, min energy consumption\\
\cline{2-7}
\hline

\parbox[t]{2mm}{\multirow{9}{*}{\rotatebox[origin=c]{90}{Miscellaneous}}}
& \cite{gomez2019intelligent} & QL & Bottleneck & Current congestion, predicted future cogestion & CoDel/FQ-CoDel target delay & Max throughput, min RTT\\
\cline{2-7}
& \cite{gomez2021federated} & QL & Inter-domain network & Current congestion, predicted future cogestion & FQ-CoDel target delay & Max throughput, min RTT \\
\cline{2-7}
& \cite{de2024desired} & DQL & P4-supported network (DASH) & In-band Network Telemetry metadata (12 fields)& iRED target delay & Max QoS \\
\cline{2-7}
\hline
\end{tabular}
\end{table*}

\subsection{Buffer Size Control}
An alternative form of queue management is to dynamically adjust buffer size rather than relying on probabilistic packet dropping. These algorithms can be viewed as a specialized subset of RED, wherein the minimum threshold is always kept equal to the maximum threshold. The exact length of an optimal buffer has been a long-debated topic within the field of computer networking \cite{vishwanath2009perspectives}, as it has been shown to drastically impact performance \cite{appenzeller2004sizing}. The algorithms in this domain strive to adaptively change this length to best mitigate bufferbloat.

LearnQueue \cite{bouacida2018practical} uses Q-learning to modify the buffer size by incrementally increasing or decreasing it by the size of one packet. The observation at each timestep consists of only the current buffer size, and the reward function encourages low delay and packet loss. LearnQueue was tested in a real-world wireless setting where multiple WARP v3 boards were placed across an office room. One board serves as the access point, while the rest are senders. The results demonstrate that LearnQueue maintains high throughput levels while reducing queuing latency compared with traditional controllers such as PIE. Despite these promising results, the potential for greater optimization exists by augmenting the MDP formulation. For example, extending the observation space to include other relevant dimensions, such as throughput and queueing rates, would better contextualize the decisions made by the RL agent. Moreover, the limitations in action space are such that it cannot swiftly change its queue length and is even incapable of maintaining its current buffer size at the same level.

BO-RL (Buffer Optimization-Reinforcement Learning) \cite{sneha2021bo} marks a significant enhancement compared to the earlier LearnQueue approach. First, it adopts a considerably larger state space that is fed into a more powerful DQL algorithm. Additionally, in contrast to its predecessor, BO-RL includes a ``maintain" action by which the agent keeps the present buffer size unchanged. For testing purposes, the authors utilized \textit{ns3-gym} \cite{gawlowicz2018ns3}, a middleware that connects the C-based \textit{ns3} library to the Python-based \textit{Gym} library. However, BO-RL exhibits certain limitations, as is also acknowledged by its authors. A notable constraint is its reliance on explicit end-user feedback as part of the agent's observation (average throughput and average per-packet delay from destination nodes). While This feedback is readily available in simulation environments, it is highly impractical within real-world networks.

RLRBM (Radio Link Buffer Management) \cite{ma2022rlrbm} addresses the difficulties associated with alleviating traffic congestion and bufferbloat in 5G and beyond systems, focusing specifically on the management of buffers within Radio Access Networks (RANs). Effective AQM for base stations is a topic worth investigating, as they can easily become transmission bottlenecks given the massive number of connected devices. The state representation merges familiar metrics, such as congestion levels, data arrival rates, and packet drop counts, with unique parameters like the downlink channel capacity and the RLC-transmitted buffer. The state also considers the ten most recent observations, which are given to an A3C algorithm. The action is defined as the maximum size of the buffer in terms of bytes, enabling the agent to flexibly modify the queue length to any size within a single timestep. RLRBM has been evaluated in a simulated urban scenario in which a walking UE moves around obstacles while sending messages to the base station. The achieved throughput of RLRBM outperforms CoDel, proving its effectiveness.

DeepAAQM \cite{song2024deep} introduces a novel approach for queue management in IoT wireless sensor networks (WSN). The paper's main goal is to balance Age of Information (AoI) and power consumption in WSNs. AoI is a measure of the freshness of information and is a crucial metric for many monitoring-based IoT applications \cite{abd2019role}. This paper's case study network is a clustered WSN where sensors relate data to the cluster heads that are in direct communication with a base station \cite{afsar2014clustering}. The RL algorithm lies in the cluster heads, where it must decide whether to flush, retain, or forward the buffered data to the base station. DeepAAQM was tested on two commonly used channel access protocols: slotted ALOHA and CSMA/CA. It showed significant reductions in power consumption while maintaining acceptable AoI levels. This research further showcases that RL, unlike traditional AQM, can be easily modified to optimize any desirable objective.

\subsection{Miscellaneous}
\cite{gomez2019intelligent} adapts a unique mixed-learning strategy where a supervised model trained to predict network congestion provides its output as part of a reinforcement learning agent's observation. Here, only the workings of the dynamic parameter tuner will be discussed, as the details of the supervised prediction mechanism are covered in the next section. Unlike previous works where the action consisted of either determining the drop/mark probability or specifying the optimal queue length, in this work, the agent controls the queue by tuning the target delay of a CoDel or FQ-CoDel algorithm. This parameter represents the desired average delay that packets experience in the queue, and it is conventionally set at a static value of a few milliseconds. The authors design the reward to be equal to that of the power function of the connection, which is defined as the ratio of throughput to RTT. After experimentation on a bottleneck network, the results showed that the intelligent RL-fitted FQ-CoDel model achieved considerably higher cumulative power function levels.

As a continuation of their previous work, the same authors introduced FIAQM \cite{gomez2021federated}. While the overall joint supervised and reinforcement learning architecture is similar, the goal of this model is to optimize inter-domain connections through communication at border routers. The primary innovation of FIAQM is its use of Federated Learning (FL), which enhances the training of the congestion predictor model in scenarios involving entities that are reluctant to share data; this is elaborated upon in the supervised learning section. Regarding the RL model, it is largely similar to \cite{gomez2019intelligent} as it uses the current and predectied congestion levels to tune the target delay of an FQ-CoDel model with the goal of increasing the power function of the connection.

In \cite{de2024desired}, the authors focus on enhancing the QoS of MPEG-DASH streaming services within a programmable data plane network, which allows for customization of router protocols with a specific language, such as p4. The resulting algorithm, named DESiRED, aims to tune the target delay parameter of a prior AQM algorithm named iRED \cite{de2022ired}, using In-band Network Telemetry (INT) metadata. Interestingly, DESiRED only drops packets at the ingress queue, while practically every other AQM system performs this action at the egress queue of a router. The authors report an impressive 42x increase in high-resolution video playback quality and a 90x reduction in video stalling, hinting that ML-based AQM might have more potential in enhancing video streaming protocols than the more widely tested TCP-based communications.

\section{Supervised \& Unsupervised Learning for AQM}

SL applications to AQM are more diverse compared with RL. However, the number of papers is comparatively limited. This may be because fitting an MDP to the AQM problem offers a more direct solution, while SL methods generally address only a specific part of the entire system. Moreover, research on related USL articles for AQM is even more lacking, so we have decided to include them within this section. As specified in tier 2 of Figure \ref{fig_aqmtax}, the applications of SL and USL can be divided into policy design (in the form of controller tuning), congestion prediction, and flow classification. Table \ref{table_sl-aqm} provides a summary of these methods.

\subsection{Controller Tuning}
Presented in \cite{sun2006neuron}, Neuron-PID is among the first methods to equip a PID controller with a neural network. A typical PID controller has three ``gain" parameters that must be tuned manually for acceptable performance. By adapting a neuron to adaptively modify these parameters, the fixed coefficients limitations of a PID are alleviated. In Neuron-PID, the neuron weights are modified via a supervised variant of Hebbian learning \cite{du2004pid}. This strategy integrates a teacher's signal to provide feedback on the neuron's current performance. Neuron-PID's capabilities were tested on a bottleneck link and reported to exceed five other models in managing the queue length at the predefined values.

AN-AQM (Adaptive Neuron AQM) \cite{sun2007adaptive} is an extended version of Neuron-PID meant to incorporate sending rate mismatch data alongside queue length error. Therefore, the neuron has three additional inputs concerning this metric. In \cite{li2014comparative}, a comprehensive comparison of AN-AQM and Neuron-PID is performed.

FFNN-AQM \cite{bisoy2018aqm} advances on the previous approaches by using a more powerful MLP controller. This model is given the current and reference queue lengths as inputs and outputs the drop probability. FFNN-AQM is thoroughly tested on single and muti-bottleneck simulations and proves to quickly converge around the desired target with only slight fluctuations.



\subsection{Congestion Prediction}
Congestion prediction is a task naturally suited for SL. In this context, ``congestion" refers to any measurement denoting the amount of traffic present in the network. Congestion prediction can take the form of predicting future queue lengths, input rates, or drop rates. However, the procedure for all these variables is relatively similar as they use the previous values of these metrics to predict the successive ones.

NN-RED (Neural Network RED) \cite{hariri2007nn} posits a refined RED scheme that considers the future value of the queue size, instead of the present size, for dropping packets. To obtain accurate predictions, NN-RED incorporates a singular perceptron that is fed the previous $L$ queue sizes to output its next value. As this predicted value exceeds the maximum threshold, packets begin getting dropped. Despite leveraging a rudimentary function approximator, NN-RED attains better queuing delay and stability than its predecessor, proving the viability of congestion forecasting for queue management. However, the true potential of this method would only be realized after the development of more robust ML function approximators.

$\alpha$\_SNFAQM \cite{zhani2007alpha_} compares a neuro-fuzzy model called $\alpha$\_SNF \cite{rouai2001new} and a Linear Minimum Mean Square Error (LMMSE) model to distinguish between ``light" and ``severe" congestion. The rest of the algorithm uses this information to manage the queue based on a couple of predefined rules. The neuro-fuzzy algorithm was found to produce more accurate outputs than the LMMSE model while having the advantage of being computationally less expensive.

In \cite{hu2019nonlinear}, a PID controller fused with an LSTM model, aptly named ''LSTM-PID" is developed. The rationale behind this technique is to combat the inherent lag a congestion control system faces due to delayed feedback. Basically, the varying delay resulting from large, dynamic computer networks undermines a typical PID controller's ability to efficiently manage the queue. This is where the LSTM comes into the picture. Its goal is to learn the dynamic behavior of the nonlinear TCP system to provide the PID controller with the most accurate network conditions.

As partially discussed in the previous section, \cite{gomez2019intelligent} does not directly alter the queue based on its predicted values; instead, it relays this information to the RL agent. This approach formulates prediction as a time-series problem and accordingly employs an LSTM model trained on queue input rates to estimate future congestion. The training process can be broken down into two distinct phases. Initially, the model undergoes training using an open dataset sourced from a backbone ISP internet link offered by the Center for Applied Internet Data Analysis (CAIDA). After completing 100 epochs, the model achieves a Mean Absolute Error (MAE) of 0.04 on the test data. Subsequently, the model is transferred to the emulated network environment, where it continues to learn by incorporating updates, allowing the predictor to refine its accuracy based on the evolving network conditions. This two-phase learning approach enables the model to gain a comprehensive understanding of the data during the initial stage while still allowing for fine-tuning in response to specific environmental changes, making it an attractive solution for real-world deployment.

In their subsequent work, \cite{gomez2021federated}, the authors shift their focus from single-domain communications to inter-domain ones. In this network scenario, entities known as Autonomous Systems (ASes)--usually large ISPs and content providers--exchange packets via an Internet Exchange Point (IXP). ASes generally share a somewhat complicated relationship: while they recognize the benefit of collaborating to boost their own network performance, as external congestion greatly impact their own domains, they are often reluctant to share the sensitive information needed to achieve this end with their potential rivals. As a solution to this complex challenge, the authors introduce the FIAQM architecture, which utilizes Federated Learning principles. In this setting, each learning round begins with a Learning Orchestrator transmitting the current global model to all participating routers. Each router then begins to further train the model using its unique local dataset. Afterward, the Learning Orchestrator collects the locally trained models from each router, and integrates their insights into a new improved global model, which will be sent out during the next learning cycle. By applying this federated approach, competing ASes are able to enhance their predictive models collaboratively by only sharing the learned model weights while preserving the confidentiality of their raw data. The empirical results presented in the paper indicate that this method significantly accelerates the learning process compared to traditional non-cooperative strategies.

\subsection{Flow Classification}
Flow classification plays a crucial role as a core component of an AQM system. While it cannot perform queue management on its own, by isolating similar flows into dedicated queues, it can provide the router with important information that directly enhances AQM potency. As stated in Section IV, flow classification is feasible with both SL and USL.

Introduced in \cite{majidi2020dc}, DC-ECN (Datacenter-ECN) is a multi-queueing scheme designed for DCNs to achieve L4S. DC-ECN consists of three elements. The first piece is an ML-based classifier for detecting elephant and mice flows. This classifier uses a Gaussian Process Regression (GPR) model trained on extracted features (e.g., packet size, source/destination IP, and protocol) to predict flow type. These segregated flows are subsequently fed into the second component, a dual couple queue, which manages ECN marking for each independent queue via a predefined equation. Finally, a scheduler monitors the condition of the various buffers to mitigate excessive queue buildup. To this end, it can adjust the maximum threshold of the elephant queues or even redirect a portion of mice packets to the elephant queue in case of traffic bursts.

Unlike the previous works that employed SL, \cite{latre2013cognitive} approaches the flow classification problem via a USL framework. The study was motivated by RFC 6077 \cite{rfc6077}, which identifies misbehaving senders and receivers as one of the seven main challenges in congestion control. Therefore, the algorithm's goal is to differentiate between responsive and non-responsive TCP flows as a means to restore fairness by penalizing misbehaving streams. The flow detection system is broken into four steps. First, a sextuple of statistical data (provided in Table \ref{table_sl-aqm}) is extracted from each flow. Next, any outlier flows are identified and removed. In the third stage, a DBScan algorithm \cite{ester1996density} clusters the dataset into groups with matching characteristics. Ultimately, these profiles are labeled as misbehaving, well-behaving, or unknown based on a hand-crafted algorithm. This unsupervised approach shows impressive results as it has a precision above 98\% and substantially helps the overall AQM system maintain high performance, even with a high-level presence of unresponsive flows.

\begin{table*}[!t]
\caption{List of works applying SL and USL to AQM}
\label{table_sl-aqm}
\centering
\def\arraystretch{2.3}
\hspace*{-1cm}
\begin{tabular}{| c |l| P{1.75cm} | P{2cm}  | P{5.5cm}  | P{5.25cm} |}
\hline
\rowcolor{gray}
\bf{Type} & \multicolumn{1}{c|}{\bf{\textsc{Ref.}}} & \multicolumn{1}{c|}{\bf{\textsc{Algorithm}}} & \multicolumn{1}{c|}{\bf{\textsc{Network}}} & \multicolumn{1}{c|}{\bf{\textsc{Model Input}}} & \multicolumn{1}{c|}{\bf{\textsc{Model Output}}}\\
\hline
\hline
\parbox[t]{2mm}{\multirow{6.5}{*}{\rotatebox[origin=c]{90}{Controller Tuning}}}
& \cite{sun2006neuron} & Perceptron & Bottleneck & Proportional component, integral component, derivative component & Drop probability\\
\cline{2-6}
& \cite{sun2007adaptive} & Perceptron & Bottleneck, multi-bottleneck & Proportional component, integral component, derivative component, normalized sending rate mismatch & Drop probability\\
\cline{2-6}
& \cite{cho2008adaptive} & RNN & Bottleneck & Error of current queue length compared to reference length& Optimal queue length\\
\cline{2-6}
& \cite{bisoy2018aqm} & MLP & Bottleneck, multi-bottleneck & Reference queue length, current queue length& Drop probability\\
\cline{2-6}
\hline

\parbox[t]{2mm}{\multirow{5.2}{*}{\rotatebox[origin=c]{90}{Congestion Prediction}}}

& \cite{hariri2007nn} & Perceptron & Bottleneck & Previous $L$ queue lengths & Estimated next queue length\\
\cline{2-6}
& \cite{zhani2007alpha_} & Nuerrofuzzy & Bottleneck & Input rates of previous timesteps & Estimated next input rate \\
\cline{2-6}
& \cite{hu2019nonlinear} & LSTM & Bottleneck & Current and previous queue length & Estimated next queue length\\
\cline{2-6}
& \cite{gomez2019intelligent} & LSTM & Bottleneck & Input rate of previous 10 timesteps & Estimated next input rate\\
\cline{2-6}
& \cite{gomez2021federated} & LSTM & Inter-domain network & Drop rate of previous 10 timesteps & Estimated next drop rate\\
\cline{2-6}
\hline

\parbox[t]{2mm}{\multirow{3.25}{*}{\rotatebox[origin=c]{90}{Flow Classification}}}
& \cite{majidi2020dc} & GPR & DCN & Source IP, destination IP, source port, destination port, transmission protocol, packet size & Mice/elephant flow classification\\
\cline{2-6}
& \cite{latre2013cognitive}& DBScan & Bottleneck & RTT, connection duration, CWR ratio, ECE ratio & misbehaving, well-behaving, or unknown flow classification\\
\hline
\end{tabular}
\end{table*}

\section{Challenges \& Future Research Directions}
The different methods surveyed in this paper showcase the potential for widespread deployment of ML-based AQM systems throughout the Internet. However, before this can turn into reality, several existing challenges must be addressed. Furthermore, promising new research directions need also be explored. This section outlines the most prominent of these issues and ideas.

\subsection{Challenges}
\subsubsection{Computational Overhead}
Modern ML abundantly makes use of hefty DNNs. For example, TD3, a popular actor-critic model, embodies a total of six DNNs under the hood \cite{fujimoto2018addressing}. While running these algorithms on purposefully designed hardware consisting of massively parallel GPUs can be efficient, it remains a severe open question as to what extent a typical router can consistently train and infer from these models locally, considering how their hardware has primarily been designed for other tasks. Initializing routers with pre-trained models removes the computational overhead of the training process. However, fine-tuning the weights is often necessary to fully adapt to specific hardware and network conditions. Alternatively, an SDN architecture could be deployed to manage ML-based AQM policies. In this mode, the centralized SDN controller can use resource-rich servers to train the models and solely send the weights to routers. Nonetheless, additional research and engineering will be needed to analyze and overcome this obstacle.

\subsubsection{Sample Efficiency}
Many of the documented RL methods for ``Policy Design" required hundreds of episodes to begin performing better than traditional AQM algorithms \cite{song2024deep}. Moreover, in the early iterations of an RL algorithm, the actions taken are practically random. While this poses no problem for a network simulation, in a real-world computer network, this initial subpar performance is generally unacceptable. A few potential solutions exist. One is the application of sample-efficient model-based RL algorithms \cite{moerland2023model}. Another appealing example is to perform an initial round of purely offline training with historically gathered data. This approach has seen limited use in the context of DCNs \cite{yan2021acc}, yet the potential to leverage similar information generated from a diverse network remains an unsolved challenge.

\subsubsection{Interpretability}
Most ML models, particularly DL approaches, are considered "black boxes," as it is often unclear how they reach certain conclusions. Undoubtedly, this muddled property increases the hesitations of network administrators to replace the classic interpretable AQM algorithms with ML approaches that can have unexpected, yet to be discovered behaviors. In recent years, Explainable AI (XAI) \cite{dwivedi2023explainable} has garnered interest with the hopes of better understanding the interworkings of these models. However, a comprehensive strategy to fully interpret complex model behavior has yet to be developed.

\subsection{Future Research Directions}
\subsubsection{Transfer Learning}
The knowledge acquired by a DNN resides within its learned weights. Therefore, a model trained on a specific task can serve as the foundation for a similar task. This whole process is known as transfer learning, and it has become popular in recent years as it universally applies to all ML subsets \cite{weiss2016survey}. For the task of AQM, investigating the extent to which experience gained from one router can be transferred to another can provide invaluable information. If transfer learning for AQM is viable, it can substantially mitigate the problem of sample efficiency stated above. To the author's knowledge, no existing paper explores transfer learning in the context of AQM.

\subsubsection{Smart Queue Management}
AQM is just one of the many mechanisms used by routers to ensure high network performance. Packet scheduling, traffic shaping, rate limiting, and QoS are examples of other data plane functions. Historically, these procedures have been thought of as distinct components, and thus,  algorithms have been developed separately without contemplating their effect on the larger ecosystem. Smart Queue Management (SQM) represents a refined router design paradigm that advocates for the integration of various functions into a unified package \cite{hoiland2018piece}. It will be interesting to see how ML can fit into this larger system. A limited number of approaches have studied the interactions of an RL-based weighted-fair queueing system with an AQM algorithm \cite{forero2021active, fawaz2021deep}. However, a comprehensive SQM system utilizing ML for multiple router functions has yet to be developed.

\subsubsection{Real-world Benchmarking}
As noted in RFC 7928, while the bottleneck topology serves as a good initial testbed for comparing algorithm performance, it does not perfectly reflect actual topologies. Despite this fact, almost all works test only on this one scenario. While considering more diverse computer simulations could provide more accurate results, the best-case scenario is to test in the real world. For instance, in \cite{barczyk2022aqm}, the researchers tested various conventional AQM algorithms on a university network and came to conclusions differing from those drawn previously from simulations. A similar test incorporating and comparing ML-AQM algorithms can have major benefits for the field and will certianly uncover never-before-seen challenges. Additionally, the accumulation of real-world data can facilitate the development of highly detailed Digital Twin Networks (DTN) \cite{wu2021digital}, which provide realistic yet risk-free simulation environments.

\subsubsection{Integration of Multiple ML Approaches}
Most ML-based AQM systems use the capabilities of one ML subset. However, many successful ML applications, such as Large Language Models (LLMs), use a mixture of all three paradigms \cite{zhao2023survey}. Therefore, it will be worth investigating how blending ML approaches can improve performance. As an example, a router can employ USL flow classification to categorize traffic into distinct queues, which are then intelligently managed by RL algorithms that also leverage future network congestion predictions made from an SL model.

\subsubsection{Testing Modern Algorithms}
When it comes to other computer networking applications, the utilization of ML-based AQM approaches still remains relatively limited. As a natural result, only a handful of ML algorithms have been explored in the context of enhancing AQM schemes. For instance, in the area of RL methods, there is a notable absence of papers using any form of model-based \cite{moerland2023model} or meta-RL \cite{beck2023survey} approaches. Additionally, in the domain of congestion prediction, many of the most well-known and advanced time series forecasting models \cite{lim2021time} have yet to be tested in any capacity.

\subsubsection{Applications to Other Networks}
Most of the discussed algorithms target wired networks, with a few addressing DCNs, and only a handful targeting some form of wireless network. However, the potential role that AQM may play in emerging network types, such as Named Data Networking (NDN) \cite{saxena2016named}, has attracted little to no attention. This also includes areas such as vehicular ad hoc \cite{rehman2013vehicular}, satellite \cite{al2022next}, and industrial networks \cite{easton2016industrial} among others. 

\subsubsection{Multi-Agent Algorithms}
The Internet is inherently a multi-agent environment, as billions of clients, routers, and servers under the orchestration of various companies and governments engage in a mixed cooperative and competitive setting.
Adding to these complications is the varying willingness of these agents to share data, as some are willing to collaborate while others are more protective of their information. The latter makes federated learning approaches appealing. Despite all of this, very few multi-agent strategies exist in the literature, and there remains a notable need for further research on how the presence of multiple intelligent agents will affect network dynamics and performance.

\section{Conclusions}
With the continued rise of novel network architectures, applications, increased loads, and higher performance standards, the need for CC persists as one of the forefront networking issues. By bringing CC to the network layer, AQM can effectively reduce latency, provided it continuously adapts to changing network conditions. The concurrent rise of ML and DL has inspired scholars to study intelligent AQM algorithms, some of which have proven initial success in both simulated and real-world environments. However, despite these promising results, considerable research and engineering efforts remain necessary for the wider adoption of such algorithms.

This paper presented the first comprehensive review of learning-based AQM algorithms. It commenced with covering the underlying principles of both AQM and ML before introducing an overview of these methods in the form of a novel taxonomy that separated algorithms based on the employed ML subset, overall objective, and implementation details. Afterward, a detailed survey of related reinforcement learning approaches was conducted, following which was an examination of applied supervised and unsupervised learning techniques. The article culminates with presenting several unresolved challenges in addition to highlighting underexplored topics suited for future research.

\bibliographystyle{plain} 


\end{document}